\documentclass[sigconf,nonacm]{acmart}

\AtBeginDocument{%
  }

\usepackage{xcolor}
\usepackage{multirow}
\usepackage{soul}
\usepackage[subtle]{savetrees}

\newif\ifcomment
\commentfalse
\ifcomment
\newcommand{\sayak}[1]{{\bf \textcolor{blue}{Sayak: #1}}}
\newcommand{\shirin}[1]{{\bf \textcolor{purple}{Shirin: #1}}}

\else
\newcommand{\shirin}[1]{}
\newcommand{\sayak}[1]{}
\fi
\begin{document}


\title{PhishLang: A Real-Time, Fully Client-Side Phishing Detection Framework Using MobileBERT}

\author{Sayak Saha Roy}
\affiliation{%
  \institution{The University of Texas at Arlington}
  \city{Arlington}
  \state{Texas}
  \country{USA}
}
\email{sayak.saharoy@mavs.uta.edu}
\author{Shirin Nilizadeh}
\affiliation{%
  \institution{The University of Texas at Arlington}
  \city{Arlington}
  \state{Texas}
  \country{USA}
}
\email{shirin.nilizadeh@uta.edu}

\begin{abstract}
In this paper, we introduce PhishLang, the first fully client-side anti-phishing framework built on a lightweight ensemble framework that utilizes advanced language models to analyze the contextual features of a website’s source code and URL. Unlike traditional heuristic or machine learning approaches that rely on static features and struggle to adapt to evolving threats, or deep learning models that are computationally intensive - our approach utilizes MobileBERT, a fast and memory-efficient variant of the BERT architecture, to capture nuanced features indicative of phishing attacks. To further enhance detection accuracy, PhishLang employs a multi-modal ensemble approach, combining both the URL and Source detection models. This architecture ensures robustness by allowing one model to compensate for scenarios where the other may fail, or if both models provide ambiguous inferences. As a result, PhishLang excels at detecting both regular and evasive phishing threats, including zero-day attacks, outperforming popular anti-phishing tools, while operating without relying on external blocklists and safeguards user privacy by ensuring that browser history remains entirely local and unshared. We release PhishLang as a Chromium browser extension and also open-source the framework to aid the research community.

\end{abstract}




\maketitle
\section{Introduction}
Phishing attacks have led to numerous data breaches and credential theft incidents, resulting in financial losses surpassing \$52 million and affecting over 300,000 users in the past year alone~\cite{ciscotrend,cnbc100million:2019}. These scams are particularly effective due to sophisticated social engineering techniques~\cite{downs2007behavioral} that allow attackers to convincingly imitate legitimate websites. In response, researchers have focused on developing countermeasures that leverage distinctive phishing indicators, such as URL characteristics, network behavior, heuristic features in source code, and advanced machine learning techniques~\cite{nguyen2013detecting,sreedharan2016systems}. Recently, there has been a growing emphasis on the visual characteristics of phishing sites~\cite{akhawe2013alice,mohammad2015phishing}, leading to the use of image-based methods, such as screenshots, for phishing detection~\cite{abdelnabi2020visualphishnet,lin2021phishpedia}.
Despite achieving high detection rates in controlled environments, these models often encounter challenges in real-world applications due to sophisticated evasion tactics used by attackers~\cite{apruzzese2022spacephish,basit2021comprehensive}. 

Keeping these models updated presents significant challenges, as attackers continuously develop new evasion tactics. Updating models to counter these threats often necessitates creating revised ground truth datasets and identifying novel feature sets or strategies. Additionally, the inherent complexity and resource demands of recent deep learning models—especially those that rely on image content—limit their practicality for large-scale, real-time deployment~\cite{saha2023phishing,oest2020sunrise}. Furthermore, the reliance on opaque, black-box approaches complicates explainability~\cite{do2022deep}, hindering opportunities for iterative improvement.

To address these issues, we introduce PhishLang, an efficient and transparent phishing detection framework utilizing MobileBERT, a lightweight language model. PhishLang employs advanced linguistic capabilities to identify subtle phishing patterns in the \emph{website source code}, thus moving beyond static feature sets. Our model operates with significantly lower resource requirements—using up to almost seven times less memory than the next resource efficient model (DistilBERT), while outperforming several ML-based models in detecting new attacks, particularly those with evasive features. Unlike most ML-based phishing detection approaches, we also evaluate PhishLang against various realistic adversarial attacks and implement patches to enhance the model's robustness. 

To the best of our knowledge, we have also developed the first open-source, Chromium-based browser extension that \textit{operates entirely on the client side}, running locally on the user's device without relying on an online blocklist, thus enhancing user privacy and eliminating blocklist-related latency. The PhishLang extension is designed to be highly efficient, with minimal memory usage, allowing it to run smoothly even on low-end hardware while delivering near-instant URL predictions. 
Despite being underexplored, client-side anti-phishing detection offers crucial benefits in phishing detection: it enhances user data privacy by keeping browsing data local, thus minimizing exposure risks, and eliminates the reliance on often delayed online blocklist updates~\cite{oest2020phishtime}. Furthermore, client-side detection reduces network latency, providing faster threat responses, ensuring continued protection during server outages. Our implementation significantly surpasses existing client-side phishing solutions, such as Google Safe Browsing, which detects only 14\% of phishing websites without server-side support~\cite{pourmohamad2024deep}.

The structure of the paper is as follows: \textbf{Section~\ref{method}} elaborates on our methodology for parsing source code to isolate critical tags associated with phishing functionality (Section~\ref{source_code_parsing}). This technique significantly reduces the input data size for the language models, thereby improving both training and inference efficiency. We use this method to create our training ground truth from the open-source PhishPedia~\cite{lin2021phishpedia} dataset (Section~\ref{training_data}). To balance processing speed with detection accuracy, we evaluate various leading open-source language models to identify the most effective one for our needs (Section~\ref{optimum-model}).

\textbf{Section~\ref{real_time_framework}} outlines the development and implementation of our real-time phishing detection system, PhishLang, which continuously monitors live URLs from Certstream to detect new phishing threats (Section~\ref{identifying_new_urls}). To benchmark our model's effectiveness against existing measures, we also compare PhishLang’s performance with traditional deep learning-based phishing models, as well as (Section~\ref{comparison_with_other_ml}) established anti-phishing tools, with a focus on detecting evasive phishing tactics (Section~\ref{blocklist_measurement}). We then evaluate the robustness of our model against adversarial phishing websites by testing sixteen realistic attack scenarios that manipulate website source code, developing \textit{patches} as countermeasures (Section~\ref{adversarial_impact}).
Finally, in Section~\ref{client-side-extension-main}, we highlight our client-side phishing detection extension.
The primary contributions of our work are:
\begin{enumerate}
\item We utilized MobileBERT, a resource-efficient language model to build a lightweight open-source phishing detection framework - PhishLang, that can not only outperform several popular machine learning-based
models, but also provide a better trade-off between inference speed, space complexity, and performance, making it more viable for real-world implementations. 
 
\item Running PhishLang on a live URL stream of URLs from 28th September 2023 to January 11th, 2024, we identified 25,796 unique phishing websites. Lower coverage of these attacks by both blocklist and URL hosting providers, especially for evasive attacks, led to PhishLang reporting and successfully bringing down 74\% of these threats, indicating its usefulness to the anti-phishing ecosystem. 
\item We provide six countermeasures (patches) that make PhishLang very robust against highly effective and realistic adversarial attacks that make perturbations in the problem space without modifying the layout of the website.
\item We open-source our framework (\url{https://github.com/UTA-SPRLab/phishlang}) and implement PhishLang as a fully client-side Chromium browser extension with very little memory impact and can run on low-end hardware, that provides protection without online blocklists. 
\end{enumerate}
\section{Related Work} 
\textbf{Phishing detection measures:}
Numerous studies have explored the various techniques employed by attackers to execute phishing  attacks~\cite{chiew2018survey,aleroud2017phishing,gupta2018defending}. 
Insights derived from these studies have been subsequently utilized to develop various detection measures. 
Scholars have introduced mechanisms for detecting phishing websites that leverage a combination of URL-based (such as Domain reputation, SSL Certification abuse, level of domain and reputation of TLD, etc.) and source code-based heuristic features (such as DOM content, presence of form fields, encoding exploits, etc.)~\cite{nguyen2013detecting,sreedharan2016systems} as well as machine learning approaches~\cite{sahingoz2019machine}. However, several studies~\cite{tewari2016recent,oest2018inside} have highlighted that these models tend to overfit the features they are trained on and exhibit a lack of resilience against evasive attacks. 
The common factor in all these cases was the automated models were not trained on the newer features, or the features were simply not available. 
For a phishing attack to be successful, it is imperative that the message is effectively conveyed to the potential victim~\cite{akhawe2013alice}.

This intuition led to the development of frameworks that study the screenshot of the website~\cite{afroz2011phishzoo,subramani2022phishinpatterns} to identify suspicious features, comparing their appearance with that of known phishing websites~\cite{jain2017phishing}, or identifying malicious artifacts, such as brand logos~\cite{liu2022inferring}. 
While showing good performance in a controlled research environment, in practice, these models are extremely resource-intensive with respect to processing power and storage required, as well as much slower than traditional heuristic or machine learning-based phishing detection methods~\cite{oest2020phishtime,saha2023phishing}. Such limitations severely restrict their utility in real-world anti-phishing frameworks, which must process millions of URLs daily~\cite{oest2020sunrise,APWG_Trends_Report_Q4_2023}.
 \newline
\textbf{Language models in Content Moderation:} Open-source language models have previously shown promise as content moderation tools in diverse areas of social computing and security. 
For example, fine-tuned BERT models are used to identify cyberbullying instances~\cite{behzadi2021rapid} and to discern toxic triggers on Reddit platforms~\cite{chong2022understanding}. 
Demicir~\cite{demirci2022static} combined LSTM and GPT-2 techniques to spot elusive malware. 
Several of these papers have demonstrated how LLMs can be trained on small amounts of data, with minimal data processing and without the need for defining specific features, which in turn outperform several traditional machine learning-based methods. 
However, utilizing these models to identify online scams and malicious websites has been very limited. Existing work~\cite{koide2023detecting,heiding2023devising} have predominantly focused on using proprietary LLMs like ChatGPT~\cite{openaichat2023}, Claude~\cite{claude2023}, and Perplexity~\cite{googlebardchat}. These models, while powerful, possess a proprietary architecture, thus preventing potential enhancements or alterations by the broader research community.
Their reliance on a cost-per-use API model~\cite{openai-pricing} further impedes scalability— a critical component in the fight against harmful content~\cite{siad2023promise}. 
Moreover, Derner at al. ~\cite{derner2023beyond} has pointed out how attackers can adversarially exploit these commercial LLMs to evade classifications, thus significantly impacting their protective capabilities. A notable exception towards utilizing language models for scam detection is Su et al.~\cite{s23208499}, who have attempted to use BERT to detect malicious websites, though the analysis is limited only to the URL of the website. 
Thus, the current state accentuates the need for an efficient open-source model tailored for phishing website detection, which is also resilient against adversarial attacks.  
\newline

\section{Method}
\label{method}

 \begin{figure*}[]
\centerline{\includegraphics[width=0.8\textwidth]{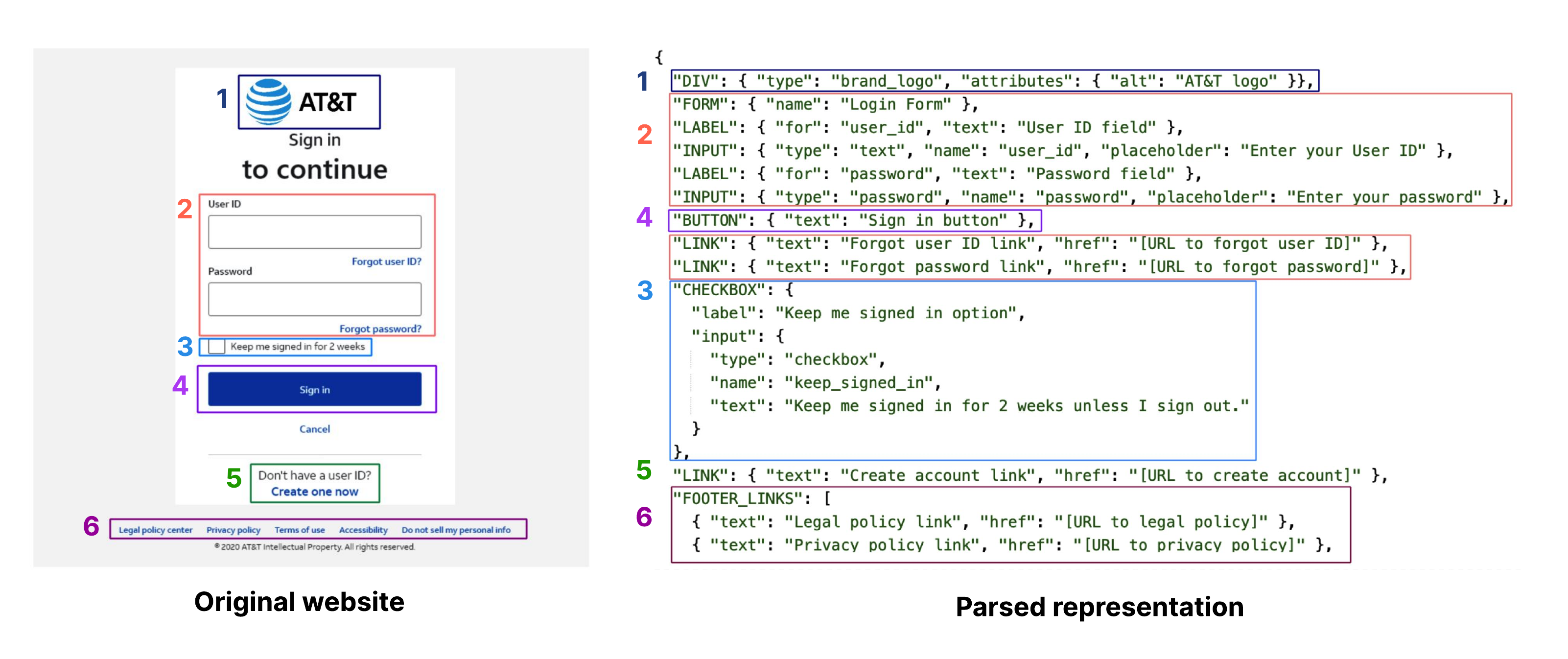}}
\caption{Example of a parsed snippet for a phishing website, with the parsed features mapped. The parsing focuses on extracting actionable features from the website. }
\label{fig:parsed_snippet}
\end{figure*}

Large Language Models excel at comprehending the semantics of natural language, enabling them to detect subtle nuances and patterns in text~\cite{zhang2020semantics,sarti2022it5}. This capability is particularly beneficial for understanding complex textual data. The flexibility of LLMs extends to the realm of source code analysis as well. Researchers have effectively utilized models like BERT to analyze source code snippets, creating embeddings that capture the contextual relationships within the code~\cite{kanade2020learning}, which have also proven useful for applications such as code completion~\cite{ciniselli2021empirical}. Building on this foundation, LLMs have been adapted for malware identification~\cite{rahali2023malbertv2,yesir2021malware}. By training on a large repository of source code, these models can predict the characteristics of the syntax that should appear in a malicious file versus that in a benign file, detecting threats without the need for specific feature definitions or predefined rules during the training stage. 
We use a similar methodology for building our phishing detection model, which is capable of anticipating the necessary content (text or syntax) in a website to classify it as phishing, both at a macro level (the overall website context) and a micro level (specific code snippets). More specifically, if any part of a website's source code - or a combination thereof—resembles that of known phishing sites, it is flagged as phishing. Conversely, if it reflects the characteristics of known benign sites, it is marked as safe. 
We break down the processes in our architecture below:

\subsection{Parsing the source code} 
\label{source_code_parsing}
A large part of a website's source code consists of aesthetic features, which include visual elements like layouts, colors, and styles~\cite{huang2013commerce} -  primarily serving to enhance the user experience but are not reliable indicators for phishing detection~\cite{alkhozae2011phishing}. Both legitimate and phishing websites may include high-quality designs~\cite{abdelnabi2020visualphishnet} (in some cases, they might even utilize identical templates~\cite{saha2023phishing}). Moreover, these aesthetic components are generally unrelated to the actionable areas of a website—such as forms, input fields, buttons, and text which users interact with, and thus where phishing tactics are typically included~\cite{chen2011interface}. 
Thus, including them in our training data not only introduces artifacts that will not be useful for phishing detection and may impact model performance but also increases the overall complexity and the time required for training the model, as the LLM needs to create more complex embeddings to map relationships between different tokens in the dataset~\cite{feng2020language}.  

Thus, to identify tags that will be relevant for phishing detection, we at first manually evaluated a set of 500 phishing websites from our training dataset (See Section~\ref{training_and_inference}), co-relating the source code of each website with its rendered website (as well as it's screenshot) to identify several tags which are used to include \textit{actionable} phishing objects, i.e., elements in the source code that are either involved in user interactions, execute some malicious functionality or contain deceptive text, that cannot be removed without affecting functionality needed by attackers need for their malicious intent (e.g., capturing credentials). We dedicate the proceeding text to listing the chosen tags and why they are relevant:
We analyze \texttt{Headings (h1, h2, h3)} as they are used to grab attention and may contain misleading language on phishing sites. \texttt{Paragraphs (p)} contain the main textual content and are often employed to convey deceptive messages. \texttt{Links (a)} are critical because they can redirect users to malicious sites or mimic legitimate URLs. \texttt{Lists (ul, ol, li)} on phishing sites might outline deceptive instructions. \texttt{Form Tags (form)} are used to collect sensitive data, making them a strong indicator of phishing.
We also use \texttt{title} tags which appear on the browser tab and attackers might insert misleading titles that closely mimic legitimate sites. For example, a phishing site might use a title such as "Secure Banking Login" to deceive users into thinking they are accessing their bank's official website, which might give the (phishing website) credibility. We also include \texttt{Footer} tags as benign websites often contain more links or information in the footer compared to phishing sites~\cite{yuan2024adversarial}, since the former is trying to provide resources to enhance user experience, whereas phishing attacks may provide little to no information in the footer.
\texttt{Script Tags (script)} are utilized for various malicious purposes, including capturing keystrokes or loading phishing content. \texttt{Input Tags (input)} in phishing sites are designed to illicitly gather personal information. \texttt{Button Tags (button)} can be manipulated to execute harmful actions such as downloading malware. \texttt{Iframe Tags (iframe)} allow the stealthy embedding of malicious content. \texttt{Meta Tags (meta)} related to redirection can automatically redirect users to fraudulent sites. \texttt{Anchor Tags with JavaScript (a with JS)} can run malicious scripts upon clicking. \texttt{Image Map Tags (map, area)} are used in phishing attempts to create misleading graphics or hidden links. We use these tags to build a parser that can automatically retrieve content enclosed within these key HTML tags, and each selected element is processed to extract its textual content and specific attributes. The output from this parsing process is formatted in a structured text representation.
Each element is represented by its tag type, followed by its content and relevant attributes. Figure~\ref{fig:parsed_snippet} illustrates a phishing website and its subsequently generated parsed representation.
Even tags that do not contain any elements are preserved with <EMPTY> tag notations. This is because phishing sites frequently utilize empty <div> or <a> tags to simulate legitimacy, misleading the users. All text content in the tags is converted to lowercase since LLMs are often sensitive to case differences, which can be interpreted as different tokens. Also, the number of unique tokens that the model needs to learn is significantly reduced.
To give further credence to our tag selection, we queried GPT-4 with 3,000 true positive samples from our training dataset (Section~\ref{training_data}), asking it to identify tags crucial for phishing detection in these websites. Prior literature has evaluated GPT-4 as highly proficient at detecting phishing websites~\cite{chataut2024can,koide2023detecting}, making it a suitable choice for our analysis. 
The model confirmed all our selected tags and additionally suggested the \texttt{applet} and \texttt{img} tags.
We chose not to include the \texttt{applet} tag, as it is outdated and deprecated in HTML5, which is predominantly used by modern phishing websites~\cite{oest2020phishtime}. Only 5 of the websites provided to GPT-4 and 42 in the entire dataset contained an \texttt{applet} tag.
Similarly, we excluded the \texttt{img} tag because it typically contains unique image filenames or externally sourced images Our framework, based on a language model, is not equipped to process images.

\subsection{Training and Inference}
\label{training_and_inference}
\subsubsection{Groundtruth}
\label{training_data}
\begin{table*}[]
\caption{Distribution of HTML tags in the dataset utilized by PhishLang. `\# of websites' denotes the number of websites at least one instance of a tag appeared in, and `Total' denotes the cumulative number of times a tag appeared in our dataset. <title> and <footer> not present in the table since they are single tags.}
\label{html_tags_dist}
\resizebox{1\textwidth}{!}{%
\begin{tabular}{c|c|c|c|c|c|c|c|c|c|c|c|c|c|c|c}
\hline
Type & Tags & \textless{}script\textgreater{} & \textless{}p\textgreater{} & \textless{}ol\textgreater{} & \textless{}h\textgreater{} tags & \textless{}a\textgreater{} & \textless{}iframe\textgreater{} & \textless{}ul\textgreater{} & \textless{}map\textgreater{} & \textless{}meta\textgreater{} & \textless{}form\textgreater{} & \textless{}area\textgreater{} & \textless{}button\textgreater{} & \textless{}li\textgreater{} & \textless{}input\textgreater{} \\ \hline 
\multirow{4}{*}{Phishing} & Min/Max & 0/239 & 0/790 & 0/22 & 0/167 & 0/636 & 0/41 & 0/153 & 0/9 & 0/324 & 0/47 & 0/48 & 0/118 & 0/698 & 0/246 \\ 
 & Mean/Median & 6.62/3.00 & 4.06/1.00 & 0.04/0.00 & 1.83/1.00 & 17.38/6.00 & 0.43/0.00 & 2.36/0.00 & 0.01/0.00 & 5.58/3.00 & 1.31/1.00 & 0.07/0.00 & 1.43/0.00 & 11.03/0.00 & 8.50/4.00 \\ 
 & \# of websites & 24780.00 & 16958.00 & 652.00 & 14962.00 & 24789.00 & 7702.00 & 13313.00 & 269.00 & 27304.00 & 26221.00 & 269.00 & 13862.00 & 13485.00 & 26879.00 \\ 
 & Total & 192396.00 & 117861.00 & 1085.00 & 53204.00 & 504693.00 & 12557.00 & 68634.00 & 339.00 & 162104.00 & 38139.00 & 2077.00 & 41577.00 & 320300.00 & 246954.00 \\ \hline
\multirow{4}{*}{Benign} & Min/Max & 0/573 & 0/20509 & 0/190 & 0/2244 & 0/12107 & 0/52 & 0/4224 & 0/76 & 0/1895 & 0/556 & 0/1024 & 0/1007 & 0/9089 & 0/2335 \\ 
 & Mean/Median & 27.01/22.00 & 28.41/13.00 & 0.47/0.00 & 17.22/10.00 & 194.77/121.00 & 2.07/1.00 & 21.45/11.00 & 0.03/0.00 & 13.93/11.00 & 2.48/1.00 & 0.18/0.00 & 6.65/2.00 & 116.79/57.00 & 11.39/4.00 \\ 
 & \# of websites & 21678.00 & 20239.00 & 3111.00 & 20438.00 & 21720.00 & 14511.00 & 20366.00 & 285.00 & 22025.00 & 17189.00 & 278.00 & 15792.00 & 20363.00 & 18340.00 \\
 & Total & 637546.00 & 670651.00 & 11206.00 & 406449.00 & 4598235.00 & 48838.00 & 506509.00 & 655.00 & 328771.00 & 58595.00 & 4287.00 & 156978.00 & 2757207.00 & 268962.00 \\ \hline
\end{tabular}}

\end{table*}

To establish a ground-truth dataset for training our classifier,  we utilized Lin at el's PhishPedia dataset~\cite{lin2021phishpedia}, which includes 30K phishing and 30K benign websites. PhishPedia is not only the largest publicly available phishing dataset but is also notable for its reliability, having been used in several prior literature~\cite{liu2022inferring,guo2021hinphish}. Each entry in this dataset typically includes associated metadata, such as HTML source code, screenshots, and OCR text. The phishing URLs were originally obtained from OpenPhish~\cite{openphish}, a widely recognized antiphishing blocklist. These URLs underwent preprocessing to eliminate inactive sites and false positives, and the researchers also implemented strategies to counteract evasive tactics like cloaking and conducted manual verifications to correct any inaccuracies in target brand representations.
Conversely, the benign URLs were sourced from the Alexa Rankings list~\cite{alexa}, a (now defunct) online database that ranked websites based on popularity.
For the purposes of training our model, we only required the HTML source file, such that it can be parsed into a collection of relevant segments as previously detailed in Section~\ref{source_code_parsing}. Out of the 30K phishing entries, 22,419 had an HTML file, whereas the same number for benign URLs was 26K.
We further analyzed this dataset beyond targeted organizations~\cite{lin2021phishpedia}, to obtain the distribution of HTML tags, libraries and frameworks used, and JavaScript APIs and function calls. Table~\ref{html_tags_dist} illustrates the distribution of HTML tags in both our phishing and benign websites. Notably, phishing websites contain \textit{fewer} tags than benign websites. This is because phishing websites primarily focus on capturing user credentials, whereas benign websites need to offer a wide range of functionalities, requiring a more extensive and complex HTML structure. We also used Wappalyzer~\cite{wappalyzer} and Esprima~\cite{esprima} to identify the frameworks and JS functions used by and present respectively in these websites. 
Finally, we are interested in identifying whether PhishLang can detect evasive phishing attacks, although our model cannot categorize the types of evasion directly. Therefore, we applied separate heuristics, derived from specific studies~\cite{oest2020phishtime, secnhack-iframe-injection, auth0-clickjacking, liang2016cracking, cybersecurityventures_punycode_phishing}, to classify the phishing URLs in the Phishpedia dataset into five categories: i) Regular phishing attacks (18,401), ii) Behavioral JS Evasions (715), iii) Clickjacking attacks (1,726), iv) Exploiting DOM (536), and v) Text encoding attacks (1,041). Behavioral JS Evasions, as highlighted by Oest et al.\cite{oest2020phishtime}, include JS-based redirection triggered by user interactions, mouse movement detections, and popup interactions, exploiting weaknesses in several prominent blocklists. In contrast, DOM Manipulations involve altering the HTML Document Object Model (DOM) to evade detection, such as dynamically generating phishing content, hiding malicious elements with CSS, and modifying the DOM structure post-page load. Liang et al.~\cite{liang2016cracking} identified features of these tactics that bypass Google's phishing page filter (GPPF), which are also transferable to other detection models. Text encoding attacks include obfuscating the character encoding, nested encoding, encoding to obfuscate JavaScript (JS), and white space obfuscation, hiding malicious code within seemingly harmless text by manipulating how the text is interpreted by different browsers. We use 70\% of our ground truth data for training the model and reserve 30\% for testing. The model is trained using 5-fold cross-validation, with each fold undergoing 10 epochs of training.

\subsubsection{Choosing the optimum model}
\label{optimum-model}
To choose the language model most suitable for building our classifier, we evaluated and compared nine language models that are considered to provide good performance in binary classification tasks (the number of parameters for the models in parenthesis): DistilBERT(66M)~\cite{sanh2019distilbert}, TinyBERT(15M)~\cite{jiao2019tinybert}, DeBERTA-base~\cite{he2020deberta}, MobileBERT (25M)~\cite{sun2020mobilebert}, FastText (10M)~\cite{grave2018learning}, and GPT-2 (117M)~\cite{radford2019language}, Llama (7B), Bloom and T5. Our aim was to choose the model which provides the best trade-off between performance, speed, and memory usage. We trained each of these models on the training split (70\%) of our ground truth dataset and tested them on our test split (30\%). 

\begin{table}[]
\caption{Performance of various language models. Time and Memory denote median time and Memory (DRAM/VRAM based on model) required for inference.
}
\label{tab:model-comparison}
\resizebox{1\columnwidth}{!}{%
\begin{tabular}{l|c|c|c|c|c|c}
\hline
\textbf{Model} & \textbf{Accuracy} & \textbf{Precision} & \textbf{Recall} & \textbf{F1 Score} & \textbf{Time (s)} & \textbf{Memory} \\ \hline
\textbf{MobileBERT} & \textbf{0.96} & \textbf{0.95} & \textbf{0.96} & \textbf{0.96} & \textbf{0.39} & \textbf{74MB} \\ 
DistilBERT & 0.94 & 0.94 & 0.94 & 0.94 & 0.85 & 502MB \\ 
DeBERTa & 0.83 & 0.83 & 0.84 & 0.84 & 1.02 & 1,341MB \\ 
FastText & 0.62 & 0.65 & 0.63 & 0.63 & 1.43 & 201MB \\ 
GPT-2 & 0.68 & 0.64 & 0.69 & 0.68 & 1.81 & 922MB \\
TinyBERT & 0.85 & 0.88 & 0.82 & 0.84 & 0.78 & 495MB \\ 
Llama2 (7B) & 0.97 & 0.96 & 0.96 & 0.96 & 33.71 & 4,873MB \\ 
T5-base & 0.89 & 0.88 & 0.88 & 0.88 & 12.40 & 1,279MB \\
Bloom (560M) & 0.75 & 0.74 & 0.74 & 0.74 & 7.49 & 7,352MB \\ 
\hline
\end{tabular}}

\end{table}

The performance of these models is illustrated in Table~\ref{tab:model-comparison}. 
\begin{figure*}[]
\centerline{\includegraphics[width=1\textwidth]{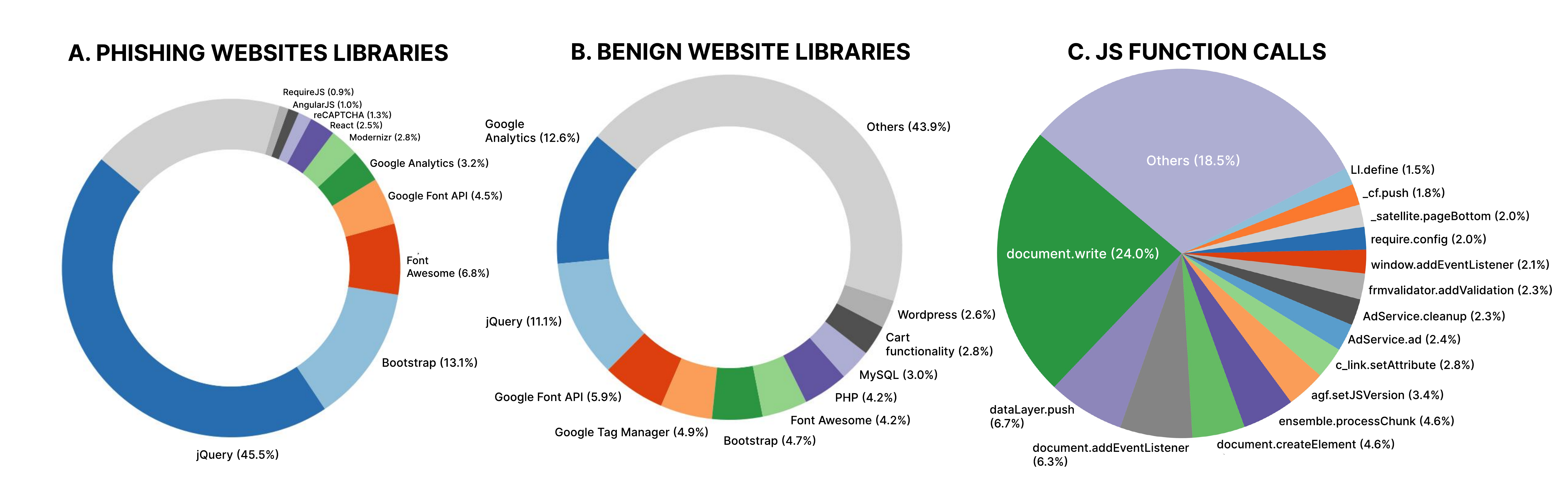}}
\caption{Distritbution of: Libraries/Frameworks found in: A) Phishing websites, B) Benign Websites, and JS Main function calls found in: C) Phishing websites.}
\label{fig:dist_libraries_jsfunctions}
\end{figure*}
We found Llama 2 to have the best performance out of all the models tested, with an F1 score of 0.96. However, its median inference time of ~34s is impractical for real-world usage where a phishing detection tool would need to process hundreds of thousands of URLs every day, and it has the highest median memory usage at 4,873.47MB. While the other two LLMs, Bloom and T5, had lower median inference times of 7.49 and 12.40 seconds respectively, they also performed worse (F1 scores of 0.88 and 0.74, respectively) and had high memory requirements (7351.89MB for Bloom and 1,279.49MB for T5). GPT-2 had the lowest median inference time (1.81 seconds) but also had the worst performance among the LLMs (F1 0.68) and required 922.41MB of memory.
Moreover, all four LLMs required inference over GPUs (while they can be used for inference using the CPU, the prediction times will be much slower), and consequently required a large amount of VRAM for said inference per sample. Requiring GPU to provide good inference time is not ideal in a practical setting, as most consumer-end devices where the inference needs to occur (such as portable/mobile systems) might not have GPUs.

When considering the Small Language Models (SLMs), MobileBERT had the highest accuracy level at 0.96, with both precision and recall scoring 0.95 and 0.96, respectively. It also demonstrated the fastest prediction times among the SLMs, with a median of 0.39 seconds per prediction, and required the least memory usage at 74MB among all models tested. These attributes make MobileBERT highly suitable for real-time phishing detection, balancing high performance with low computational resource requirements, which is ideal for deployment on devices with limited hardware capabilities, such as mobile and portable systems. Its efficient use of DRAM instead of requiring a GPU makes it even more attractive for widespread, scalable deployment.
In comparison, DistilBERT achieved an accuracy of 0.94, with both precision and recall scoring 0.94. While not the quickest, its prediction times were reasonably efficient, with a median of 0.85 seconds per prediction and significantly higher memory usage of 502MB compared to MobileBERT. DeBERTa-base, another SLM, delivered lower precision and recall, also fell short in time efficiency, and had a higher memory usage of 1,341MB. FastText, though more compact with a memory usage of 201MB, suffered from reduced accuracy (0.62) and had a median prediction time of 1.43 seconds. Meanwhile, TinyBERT, the smallest model tested, offered quicker results at 0.78 seconds but at the cost of lower performance metrics, requiring 495MB of memory. All SLMs could provide inference over CPUs, making them suitable for deployment on various platforms.
Also, due to the unavailability of commercial LLMs such as ChatGPT and Claude as local implementations, fine-tuning and testing on these models would incur significant costs for training and, especially for inference, given the millions of URLs an anti-phishing tool needs to process daily. However, we compare ChatGPT (3.5 Turbo) and GPT 4 with our model, and other ML-based phishing detection tools in Section~\ref{comparison_ml_models}. Thus, considering all the tested language models, based on a combination of high precision and recall, fast prediction times, and efficient memory usage, MobileBERT was the most suitable choice for our framework and was thus chosen as the final language model for our PhishLang framework.

\subsection{Comparison with other ML models}
\label{comparison_with_other_ml}

We compared the performance of PhishLang with other popular open-source ML-based phishing detection models. 
Two of these models rely on the visual features of the website: VisualPhishNet~\cite{abdelnabi2020visualphishnet} and PhishIntention~\cite{liu2022inferring}, one relies on both the URL string and  HTML representation of the website: StackModel~\cite{li2019stacking}, and one that relies on the semantic representation of the URL string only: URLNet~\cite{le2018urlnet}.  Each of these models was tested on the 5K samples, which we had labeled earlier. 
We also include the commercially available models of ChatGPT (GPT 3.5T and GPT 4). 
Since the focus of PhishLang is not only on prediction performance but also on the overhead of the models, which is crucial for practical implementation, we add two metrics - Median Artifact size, i.e., the amount of data (source code, URL, image, etc. based on the model) that needs to be provided to the model for prediction, and Median Prediction Time, which is the median time required for the respective model to provide the prediction label. Also, since GPT 3.5T and 4 are general LLM models that can predict any text content,  we pass both the full HTML and the parsed versions as artifacts for evaluation. Table~\ref{comparison_ml_models} illustrates the performance of each of the models.
We note that PhishIntention, despite showing the best performance among all the models, requires a relatively large artifact size of 348KB and has a longer prediction time of 10.7 seconds.
PhishLang, on the other hand, shows strong performance metrics with the advantage of a significantly smaller artifact size (7KB) and faster prediction time (0.9 seconds), indicating a balance between high performance and operational efficiency.
Visual PhishNet and StackModel show moderate performance in all metrics, with the former requiring a much larger artifact size and longer prediction time compared to StackModel. 
Finally, URLNet has the lowest performance metrics but benefits from the smallest artifact size (0.2KB) and shortest prediction time (0.7 seconds). 
In the case of the ChatGPT models, we found that GPT 3.5 was only able to process 822 phishing + 1031 benign samples when the full HTML was given, while GPT 4 was able to process 1,469 phishing + 2,147 benign samples due to the token size limit. 
We find that both models perform very well against the threats and are almost on par with both PhishLang and PhishIntention. Note that these GPT models are general models and were not specifically fine-tuned using phishing training data. Thus, it is possible that it can outperform our model if fine-tuned on appropriate data. However, restricting GPT models for commercial use and the overall expense required for training them could make them unfeasible for practical implementation.

\begin{table}[h!]
\centering
\caption{Comparison with other ML-based detection methods}
\resizebox{1\columnwidth}{!}{%
\begin{tabular}{l|cccccc}
\hline
Model & Accuracy & Precision & Recall & F1 Score & Size & Time \\ 
\hline \hline
\textbf{PhishLang} & 0.94 & 0.93 & 0.95 & 0.94 & 7KB & 0.9 \\
\hline
PhishIntention & 0.96 & 0.94 & 0.96 & 0.95 & 348KB & 10.7 \\
Visual PhishNet & 0.85 & 0.83 & 0.86 & 0.85 & 217KB & 4.9 \\
URLNet & 0.73 & 0.72 & 0.74 & 0.73 & 0.2KB & 0.7 \\
StackModel & 0.83 & 0.85 & 0.81 & 0.82 & 1KB & 1.9 \\
GPT 3.5T (HTML)\textsuperscript{*} & 0.87 & 0.88 & 0.86 & 0.87 & 42KB & 2.1 \\
GPT 4\ (HTML)\textsuperscript{*} & 0.92 & 0.93 & 0.91 & 0.92 & 42KB  & 2.8 \\
GPT 3.5T (Parsed)& 0.90 & 0.90 & 0.88 & 0.88 & 7KB & 1.6\\
GPT 4\ (Parsed) & 0.94 & 0.93 & 0.93 & 0.93 & 7KB  & 1.4\\
\hline
\end{tabular}}
\label{comparison_ml_models}
\end{table}

\section{Real-time framework}
\label{real_time_framework}



We implement our model as a framework that continuously identifies new phishing websites and reports them to various antiphishing entities, including browser protection tools, commercial blocklists, scanners, and hosting providers. 

\subsection{Identifying new URLs}
\label{identifying_new_urls}
We run our model on Certstream~\cite{certstream:2022}, a Certificate Transparency Log that streams all SSL-certified URLs. This platform is extensively utilized to detect new phishing URLs, as a vast majority of phishing sites now employ SSL certification~\cite{halfsslphishing}. Prior research has also frequently leveraged this data source~\cite{liu2022inferring,saha2023phishing}. 
If a website is flagged as benign with low confidence (<0.5), the model investigates the first five links in the website’s source to detect phishing attacks that might not be apparent on the landing page, such as hidden phishing elements or image-based links. We observed an average of 27 domains per second on Certstream, and despite occasional delays in processing due to network bottlenecks, the model provided predictions within a median time of 9 minutes. From September 28, 2023, to January 11, 2024, PhishLang scanned 42.7M domains (172M websites through links), flagging 25,796 as phishing (0.057\%).

We detected 17,396 regular, 3,159 JavaScript evasion, 3,349 Clickjacking, 1,057 DOM, and 835 Text encoding attacks. Some evasions might not have been identified and could have been classified as regular attacks. To verify the accuracy of the model, two coders, experienced in computer security concepts, manually assessed 2.5k websites flagged as phishing, ensuring a representative sample of each attack type. They first checked if the websites impersonated any of the 409 brands identified as common phishing targets by OpenPhish in August 2022~\cite{openphish_brands}, followed by examining if the sites solicited sensitive information or triggered suspicious downloads. Downloads with four or more detections on VirusTotal were deemed malicious, based on previous findings~\cite{peng2019opening}. The inter-rater reliability (Cohen’s kappa) was 0.82, indicating high agreement and any disagreements were mutually resolved. 
Ultimately, 2,388 sites were confirmed as phishing, demonstrating the model's accuracy of approximately 94\% for regular attacks and an average detection of 91\% for the four evasive categories. We break down the model's performance against each type of attack in Table~\ref{table:model_evasive_attacks}. It is evident that PhishLang is capable of detecting a range of evasive attacks.

\begin{table}[t]
\caption{Performance of PhishLang against evasive attacks}
\label{table:model_evasive_attacks}
\centering
\resizebox{0.9\columnwidth}{!}{%
\begin{tabular}{l|c|c|c|c}
\hline
\textbf{Attack Type} & \textbf{Samples} & \textbf{Accuracy} & \textbf{Precision} & \textbf{Recall} \\ \hline \hline
Regular attacks & 1,623 & 94.2\% & 94.5\% & 95.7\% \\ 
Behaviour based JS & 280 & 92.9\% & 93.3\% & 92.6\% \\
Clickjacking & 313 & 91.3\% & 92.1\% & 90.5\% \\ 
DOM Manipulations & 94 & 88.4\% & 89.0\% & 87.1\% \\ 
Text encoding  & 78 & 90.7\% & 91.2\% & 89.9\% \\ \hline
\end{tabular}}

\end{table}

\subsection{Evaluating Commercial Phishing Countermeasures}
\label{blocklist_measurement}
In this section, we evaluate the effectiveness of different phishing countermeasures against websites identified by PhishLang. These include two commercial browser protection tools—Google Safe Browsing and Microsoft SmartScreen—and two open-source phishing blocklists, PhishTank~\cite{phishtank} and OpenPhish~\cite{openphish}. Safe Browsing is the default protection in Google Chrome, Mozilla Firefox, and Safari, while SmartScreen powers Microsoft Edge, collectively protecting nearly 95\% of all internet users~\cite{statcounter_browser_share}. We assess how many phishing websites identified by PhishLang are covered by these tools and whether our reporting helps detect the initially missed websites. PhishTank and OpenPhish, community-driven blocklists, are used by several commercial antiphishing tools~\cite{phishfriends,oest2020sunrise}.
PhishLang checks each identified phishing URL against these tools and blocklists in real time. If a URL is not covered, it is reported, and we monitor its detection status for up to a week. Considering ethical implications, we report all 26k identified websites immediately to ensure user protection, despite our 97\% accuracy rate suggesting a potential 3\% false positives being reported. This trade-off is acceptable given the typically higher noise in blocklist data~\cite{saha2023phishing,roy2021evaluating,oest2020sunrise}.
\begin{table*}[]
\caption{Performance of browser protection tools, blocklists and hosting domains}
\label{browser-protection}
\centering
\resizebox{0.9\textwidth}{!}{%
\begin{tabular}{c|c|cc|cc|cc|cc|cc}
\hline
\multirow{2}{*}{Attack Type} & \multirow{2}{*}{\# URLs} & \multicolumn{2}{c|}{Safe Browsing} & \multicolumn{2}{c|}{SmartScreen} & \multicolumn{2}{c|}{PhishTank} & \multicolumn{2}{c|}{OpenPhish} & \multicolumn{2}{c}{Hosting Domain} \\ \cline{3-12} 
 &  & \multicolumn{1}{c|}{Detection (\%)} & Time & \multicolumn{1}{c|}{Detection (\%)} & Time & \multicolumn{1}{c|}{Detection (\%)} & Time & \multicolumn{1}{c|}{Detection (\%)} & Time & \multicolumn{1}{c|}{Detection (\%)} & Time \\ \hline
Regular Attack & 1,595 & \multicolumn{1}{c|}{50.92 (84.33)} & 3.71 hrs & \multicolumn{1}{c|}{34.26 (74.26)} & 5.85 hrs & \multicolumn{1}{c|}{14.8 (36.9)} & 7.33 hrs & \multicolumn{1}{c|}{22.6\% (47.1)} & 3.81 hrs & \multicolumn{1}{c|}{39.2 (61.0)} & 4.50 hrs \\ 
JS Evasion & 306 & \multicolumn{1}{c|}{35.27 (70.36)} & 5.67 hrs & \multicolumn{1}{c|}{27.48 (75.42)} & 9.66 hrs & \multicolumn{1}{c|}{3.9  (19.4)} & 11.19 hrs & \multicolumn{1}{c|}{8.5\% (27.9)} & 5.77 hrs & \multicolumn{1}{c|}{6.4\% (19.4)} & 2.15 hrs \\ 
ReCAPTCHA/QR & 91 & \multicolumn{1}{c|}{40.97 (66.07)} & 3.75 hrs & \multicolumn{1}{c|}{15.31 (81.34)} & 7.54 hrs & \multicolumn{1}{c|}{9.9  (29.7)} & 8.02 hrs & \multicolumn{1}{c|}{16.5\% (37.1)} & 4.05 hrs & \multicolumn{1}{c|}{9.9\% (29.7)} & 10.22 hrs \\ 
Clickjacking & 325 & \multicolumn{1}{c|}{61.32 (80.28)} & 2.42 hrs & \multicolumn{1}{c|}{38.48 (77.65)} & 5.3 hrs & \multicolumn{1}{c|}{7.7  (26.8)} & 14.70 hrs & \multicolumn{1}{c|}{15.7\% (36.4)} & 7.19 hrs & \multicolumn{1}{c|}{7.1\% (26.8)} & 5.73 hrs \\ 
DOM Attacks & 102 & \multicolumn{1}{c|}{55.79 (74.39)} & 3.48 hrs & \multicolumn{1}{c|}{18.27 (85.36)} & 5.56 hrs & \multicolumn{1}{c|}{10.8 (31.0)} & 11.15 hrs & \multicolumn{1}{c|}{20.6\% (40.5)} & 13.71 hrs & \multicolumn{1}{c|}{10.8\% (31.0)} & 7.19 hrs \\ 
Text encoding & 81 & \multicolumn{1}{c|}{27.97 (72.57)} & 6.45 hrs & \multicolumn{1}{c|}{20.79 (70.36)} & 2.5 hrs & \multicolumn{1}{c|}{8.6 (28.0)} & 6.42 hrs & \multicolumn{1}{c|}{16.0\% (36.7)} & 2.88 hrs & \multicolumn{1}{c|}{8.6\% (28.0)} & 5.13 hrs \\

\hline
\end{tabular}}

\end{table*}. Browser protection tools employ heuristic and ML-based models within the browser environment to detect phishing pages~\cite{liang2016cracking}. Using Selenium~\cite{selenium_python}, we automate opening each website in a browser and take screenshots to confirm detection by these tools. Table~\ref{browser-protection} illustrates the initial coverage of these tools and subsequent coverage after we had reported the URLs that they had failed to detect. 
All tools initially have low coverage, with Google Safe Browsing outperforming SmartScreen. Reporting undetected URLs significantly improved detection rates, especially for evasive phishing attacks. For instance, Google Safe Browsing's detection of JS evasion attacks increased from 35\% to 73\%, and SmartScreen's detection of DOM-based attacks rose from 18\% to 86\%. Regular phishing URL detection also improved significantly. Despite these improvements, it is important to note that these tools generally maintain high detection rates and low false positives~\cite{oest2020sunrise}. Our model thus helps close detection gaps, particularly for new and evasive attacks, and accelerates the integration of new threat intelligence.

PhishTank and OpenPhish have different responses to reported URLs. PhishTank lists reported URLs immediately, requiring community verification before labeling them as phishing or benign, and we measure the time taken between our reporting and subsequent labeling. OpenPhish verifies URLs before inclusion in the blocklist, and we measure the time until URLs appear on the blocklist. For hosting domains, we track how long it takes for reported websites to become inactive. Table~\ref{browser-protection} shows the performance of these entities, showing significant detection rate improvements following our reports, especially for evasive attacks, indicating that PhishLang enhances these countermeasures' effectiveness.

\section{Adversarial impact}
\label{adversarial_impact}

The effectiveness of Machine Learning (ML) and Deep Learning (DL) models can be significantly reduced by adversarial inputs designed to lower the model's confidence score, which can also lead to misclassification. These inputs exploit inherent vulnerabilities related to the training data, architectural design, or developmental assumptions of the model. 
For example, in the case of Machine-Learning phishing webpage detectors (ML-PWD), Panum et al.~\cite{panum2020towards} showed how image-based phishing detection models are vulnerable to FSGM pixel-level exploitation, where adversarial websites are modified to appear like benign ones. 
AlEroud et al.~\cite{desolda2023explanations} explores how models relying on URL-based features can be exploited by employing perturbations in the URI structure. 
Both image-based features and URL-based attributes are irrelevant to our model, which solely relies on a collection of tags extracted directly from the HTML source code. Consequently, the only avenue for an attacker to exploit our model is to target the specific HTML tags we analyze and alter their associated text. 

While previous literature has investigated how attackers can modify attributes in the ``Feature space''~\cite{shirazi2019adversarial,bac2021pwdgan,corona2017deltaphish,lee2020building} (i.e., when the website is processed into feature vectors specific to those that are used by the ML-PWD for prediction), recent studies have focused more on attackers introducing perturbations in the ``problem space'' (i.e., where the attacker directly modifies the webpage)~\cite{liang2016cracking,bahnsen2018deepphish,abdelnabi2020visualphishnet}. This shift is due to real-world attackers operating in the ``problem space''~\cite{pierazzi2020intriguing}, where the perturbations introduced are subject to certain physical constraints. If these constraints are not met, which can happen when perturbations are added in the feature space, there is an increased risk of generating adversarial samples that, while potentially reducing the model's efficiency\cite{tong2019improving}, may result in samples that are physically unrealizable. 

In the context of phishing websites, this often manifests as notable artifacts in the webpage layout, as demonstrated by Yuan~\cite{yuan2024adversarial} and Draganovic et al.~\cite{draganovic2023users}. 
These artifacts make the sites suspicious to users, reducing the attacks' overall effectiveness. 
Moreover, to perturb the feature space, the attacker is required to have internal access to the processing pipeline of the ML-PWD—a process that can be challenging and costly~\cite{apruzzese2022modeling}. 
Additionally, most adversarial implementations in the literature target specific ML-PWDs, but realistically, attackers do not target specific models. Instead, they prefer generalizable evasive attacks, that manipulate raw data in the problem space, making the attack evasive across different detection systems. Therefore, we focus on testing and improving the adversarial robustness of PhishLang, emphasizing the problem-space attacks. These attacks are designed to introduce perturbations into the website source code while maintaining the website's layout.

\subsection{Evaluating Problem-space evasions}

Apruzzese~\cite{apruzzese2022spacephish} and Yuan et al.~\cite{yuan2024multi} have conceptualized evasive attacks that are relevant and generalizable for creating phishing samples capable of evading the four most popular ML-PWDs. They identified 57 features that attackers commonly use to bypass ML-PWD systems. Building on this framework, Montaruli et al.~\cite{montaruli2023raze} developed a novel set of 16 evasive attacks focusing on the problem space of adversarial attacks. These attacks involve directly modifying the HTML content using fine-grained manipulations, allowing for changes to the HTML code of phishing webpages without compromising their maliciousness or visual appearance. This approach successfully preserves both the functionality and appearance of the websites.
Some examples of these attacks include \emph{InjectIntElem}, which involves injecting a specified number of (hidden) internal HTML elements (e.g., <a> tags with internal links) into the body of the webpage and \emph{ObfuscateJS}, where the JavaScript code within <script> elements of a webpage is obfuscated by encoding them into Base64 and then inserts a new script to decode and execute the original script. A full description of these attacks is provided in the author's paper. For brevity, we do not go into comprehensive details about which feature (based on Apruzzese~\cite{apruzzese2022spacephish} and Yuan et al.\cite{yuan2024multi}) these attacks target. 

Since different phishing websites can be compromised by different attacks, Montaruli et al also developed a query-efficient black-box optimization algorithm based on WAF-A-MoLE~\cite{demetrio2020waf}. This algorithm employs an iterative methodology that involves successive rounds of mutation to modify the original malicious sample, aiming to reduce the confidence score provided by ML-PWD. Attacks are carried out as Single Round (SR) or Multi Round (MR) manipulations. SR manipulations, such as UpdateHiddenDivs, UpdateHiddenButtons, UpdateHiddenInputs, and UpdateTitle, are applied once to achieve their goal of hiding or modifying specific HTML elements and do not require further changes to evade detection. In contrast, MR manipulations, including InjectIntElem and InjectExtElem, require multiple sequential applications to progressively adjust the internal-to-external link ratios and other features to effectively evade detection.
The optimizer begins by initializing the best adversarial example and score with the initial phishing webpage and its corresponding score. It then sequentially applies the SR manipulations, updating the best example and score whenever a new manipulation achieves a lower score. Afterward, the optimizer enters the MR manipulations loop, which includes a specified number of mutation rounds. In each mutation round, the algorithm generates new candidate adversarial phishing webpages from the current best example by applying one MR manipulation to each candidate. The candidate with the lowest confidence score is selected, and if this score is lower than the current best, it becomes the new best adversarial example. The number of mutation rounds, $R$, given the maximum query budget $Q$, is determined by the formula $R = \frac{Q - \#SR}{\#MR}$, where $\#SR$ and $\#MR$ represent the number of SR and MR manipulations, respectively. 
One attack \textit{InjectFakeFavicon} (an SR manipulation) is not relevant to PhishLang since our parser does not look for favicons in \texttt{<head>} tags.
We set a query budget of 35 (instead of 36 as in the original paper due to the omission of one SR manipulation) and injected each attack into each sample in our test set (6,725 URLs) to choose the best adversarial sample. 
In addition to adversarially testing PhishLang, our goal is to introduce patches that can reduce or nullify the effectiveness of these attacks. While both SR and MR manipulations cumulatively work towards creating the best-performing adversarial sample, it is essential to identify which particular attack had the most impact. This involves determining which manipulation was instrumental in significantly reducing the confidence of PhishLang when evaluating the sample. This specific manipulation can be considered the \emph{primary attack} that was crucial in achieving the desired adversarial effect.
To identify this, after each SR and MR round, we compute the difference in the \textit{adversarial advantage} for the sample during the current stage compared to the previous stage. The adversarial advantage, \(A\), is computed based on the drop in the confidence score of the model towards the phishing website during subsequent optimization stages. Therefore, if \(C_{\text{o}}\) is the original confidence score of the model in identifying the website as phishing and \(C_{\text{p\textsubscript{1}}}\) is the confidence score after the first perturbation round (SR as per the algorithm), then we define \(A\) as $A = C_{\text{o}} - C_{\text{p\textsubscript{1}}}$.

By analyzing these differences, we can identify which manipulations had the most significant impact on reducing the model's confidence, thereby determining the primary attack that achieved the desired adversarial effect. If two or more attacks demonstrate similar adversarial advantages (within one standard deviation of the mean adversarial advantage across all samples), we designate these as primary attacks \textbf{[Scenario 1]}. 
Additionally, an attack following a primary attack may negatively impact the model's performance, but the preceding primary attack might have lessened its effectiveness. Therefore, once a primary attack is identified, we remove it from the sample and rerun the optimization algorithm. If another attack then exhibits an adversarial advantage similar to the previously recognized primary attack (within one standard deviation of the adversarial advantage across all samples), it is also designated as a primary attack \textbf{[Scenario 2]}. In our dataset, we noticed that only 11.3\% of the samples met \textbf{[Criteria 1]} and 4.5\% (samples met \textbf{[Criteria 2]}, indicating that for most phishing websites, only one attack had a significant impact in adversarially enhancing it.
Table~\ref{adv_performance} presents the performance of each \textit{adversarial attack} on PhishLang, detailing the number of incorrect predictions (false negatives) and the corresponding median adversarial advantages. We observe that all problem-space-facing attacks were able to decrease the efficiency of our vanilla model initially, which was then significantly improved by applying the patches. For example, the ObfuscateJS (A8) attack had a median adversarial advantage of 0.348, which dropped to 0.028 after applying adversarial patches, and the incorrect prediction rate reduced from 20.92\% to 1.51\%. Similarly, the UpdateHiddenButtons (A12) attack showed a median advantage of 0.241 initially, which was reduced to 0.025 post-patch application, with incorrect predictions decreasing from 17.20\% to 1.07\%. 


\begin{table}[]
\caption{Impact of different adversarial attacks. Text in parentheses indicates values post adversarial patches were applied.}
\label{adv_performance}
\resizebox{0.9\columnwidth}{!}{%
\begin{tabular}{l|l|l}
\hline
\textbf{Attack} & \textbf{Median advantage} & \textbf{Incorrectly predicted} \\ \hline
InjectIntElem (A1) & 0.204 (0) & 7.99\% (0) \\
InjectIntElemFoot (A2) & 0.142 (0) & 11.24\% (0) \\
InjectIntLinkElem (A3) & 0.231 (0) & 4.43\% (0) \\
InjectExtElem (A4) & 0.285 (0) & 7.02\% (0) \\
InjectExtElemFoot (A5) & 0.194 (0) & 15.01\% (0) \\
UpdateForm (A6) & 0.271 (0) & 4.90\% (0) \\
ObfuscateExtLinks (A7) & 0.314 (0.039) & 19.01\% (1.68\%) \\
ObfuscateJS (A8) & 0.348 (0.028) & 20.92\% (1.51\%) \\
InjectFakeCopyright (A9) & 0.201 (0) & 4.10\% (0) \\
UpdateIntAnchors (A10) & 0.245 (0) & 7.92\% (0) \\
UpdateHiddenDivs (A11) & 0.173 (0) & 8.01\% (0) \\
UpdateHiddenButtons (A12) & 0.241 (0.025) & 17.20\% (1.07\%) \\
UpdateHiddenInputs (A13) & 0.403 (0.027) & 13.59\% (1.18\%) \\
UpdateTitle (A14) & 0.107 (0.005) & 11.27\% (0.63\%) \\
UpdateIFrames (A15) & 0.111 (0) & 5.42\% (0) \\ \hline
\end{tabular}}

\end{table}

\subsection{Mitigating Adversarial Attacks}

To mitigate or nullify the effectiveness of these attacks, we employ one or both of the following strategies: a) Parser Modification - Here, we identify that the attack can be sufficiently neutralized by modifying how our framework parses the raw HTML source code, eliminating the need for adversarial retraining, and b) Adversarial Training - We retrain our model using adversarial samples in cases where the respective attack is identified as the \textit{primary attack}.

\textit{Parser modifications}: Attacks InjectIntElem (A1), InjectIntElemFooter (A2), InjectIntLinkElem (A3), InjectExtElem (A4), and InjectExtElemFooter (A5) use one of four strategies to create the adversarial samples: S1, S2, S3, and S4. S1 inserts the hidden attribute into an HTML element to prevent the browser from rendering its content, adopting the hidden attribute as its default approach. S2 modifies the style attribute to ``display\:none'' to hide elements. S3 uses the <style> HTML element to achieve the same effect as S2 but through the <style> tag. S4 places HTML elements to be hidden inside the <noscript> tag, effective only if JavaScript is enabled in the victim's web browser. To counter these attacks, we propose two parser-based modifications. \textbf{Patch 1} ignores any tags with a hidden attribute or ``display: none'' in the style attribute, which reduces the effectiveness of attacks applied using S1 and S2 to 0. For tags that refer to a CSS element, the parser searches the corresponding CSS <style> sheet for display\:none attributes. If found, the initial tag is omitted from being included in the parsed representation given to the model. This constitutes \textbf{Patch 2}, which reduces the effectiveness of S3 to 0. We did not need to implement any protection for S4, as our parser does not consider <noscript> elements. Patches 1 and 2 are also transferable to Attacks InjectFakeCopyright (A9), UpdateHiddenDivs (A11), and UpdateIFrames (A15), which use similar syntax to add or modify hidden elements in the source code. On the other hand, attack UpdateForm (A6) manipulates forms on a webpage by replacing the original internal link specified in the action attribute with a different internal link that does not trigger detection features, such as ``\#!'' or ``\#none.'' To counter this attack, we propose \textbf{Patch 3}, a parsing modification that checks if the `action' attribute directs to an internal section or is a valid external link (using Python's request library~\cite{python_requests}) that nullified the attack effectiveness to 0. Finally, ObfuscateExtLinks (A7) obfuscates the external links in a webpage to evade multiple HTML features by substituting the external link with a random internal one that is not detected as suspicious by the HTML features (e.g., \#!) and creates a new <script> element that updates the value of the action attribute to the original external link when the page is loaded. While similar to A6 in terms of manipulating links to avoid detection, A7 specifically targets external links and employs JavaScript to dynamically revert the obfuscation. Regardless, Patch 3 successfully nullifies this attack and UpdateIntAnchors (A10).

\textit{Adversarial retraining}
Attack ObfuscateJS (A8) obfuscates JavaScript within \texttt{<script>} elements to evade detection features such as \texttt{HTML\_popUP}, \texttt{HTML\_rightClick}, and \texttt{HTML\_statBar}. This technique involves extracting the original JavaScript, encoding it in Base64, and replacing the original \texttt{<script>} content with new code that decodes and executes the obfuscated JavaScript. To counter this attack, we developed two patches: \textbf{Patch 4.1} converts Base64 (or any other non-UTF-8 encodings) to UTF-8 before parsing, and \textbf{Patch 4.2} retrains the model with adversarial samples where A7 was the primary attack. Although these patches did not completely nullify the attack, they significantly reduced the median adversarial advantage to 0.039 (from 0.314) and the false negative rate to 1.68\% (from 19.01\%).
Attack UpdateHiddenButtons (A12) obfuscates all disabled button elements by removing the \texttt{disabled} attribute and inserting a new \texttt{<script>} element that re-applies this attribute during rendering using the \texttt{setAttribute()} DOM method. This approach ensures that both the rendering and original behavior of the buttons are preserved. While a potential countermeasure could involve ignoring scripts with the \texttt{disabled} attribute, this could break the user experience. For example, in a web form where the ``Submit'' button is initially disabled and only enabled after all required fields are filled out, ignoring such scripts could prevent proper functionality. Therefore, to address this attack, we retrained our model with adversarial samples where A12 was the primary attack (\textbf{Patch 5}). This reduced the median adversarial advantage to 0.p025 (from 0.241) and the false negative rate to 1.07\% (from 17.20\%). Attack UpdateHiddenInputs (A13) evades detection of hidden and disabled input elements by changing the \texttt{type} attribute from ``hidden'' to ``text'' and adding the \texttt{hidden} attribute. We adversarially trained with samples that had this attack as the primary focus \textbf{(Patch 6)}, which reduced the median adversarial advantage to 0.0.27 (from 0.403) and the false negative rate to 1.18\% (from 13.59\%). For Attack UpdateTitle (A14), we use a variant of Patch 6 that prevents JS from executing when reading the \texttt{<title>} tag.

\subsection{Advesarial impact on phishing categories}
\begin{table}[]
\caption{Impact of adversarial attacks on phishing attacks by category (Test set)}
\label{adv_phish_category}
\resizebox{1\columnwidth}{!}{%
\begin{tabular}{l|cc|cc}
\hline
\multirow{2}{*}{\textbf{Attack Type}} & \multicolumn{2}{c|}{\textbf{Before intervention}} & \multicolumn{2}{c}{\textbf{After intervention}} \\ \cline{2-5} 
 & \multicolumn{1}{c|}{\begin{tabular}[c]{@{}c@{}}Median \\ Advantage\end{tabular}} & \begin{tabular}[c]{@{}c@{}}Incorrectly \\ Predicted\end{tabular} & \multicolumn{1}{c|}{\begin{tabular}[c]{@{}c@{}}Median \\ Advantage\end{tabular}} & \begin{tabular}[c]{@{}c@{}}Incorrectly \\ Predicted\end{tabular} \\ \hline
Regular attacks (5,734) & \multicolumn{1}{c|}{0.212} & 648 & \multicolumn{1}{c|}{0.030} & 37 \\ 
Behaviour based JS (198) & \multicolumn{1}{c|}{0.425} & 73 & \multicolumn{1}{c|}{0.246} & 29 \\ 
Clickjacking (485) & \multicolumn{1}{c|}{0.230} & 91 & \multicolumn{1}{c|}{0.033} & 14 \\ 
DOM Manipulations (101) & \multicolumn{1}{c|}{0.380} & 45 & \multicolumn{1}{c|}{0.021} & 3 \\ 
Text encoding (207) & \multicolumn{1}{c|}{0.316} & 42 & \multicolumn{1}{c|}{0.063} & 8 \\ \hline
\end{tabular}}

\end{table}

We also examined the impact of adversarial attacks on our test set samples based on different attack categories. Table~\ref{adv_phish_category} shows the median adversarial impact and the number of incorrectly predicted samples for the five phishing attack categories (Regular, BJE, Clickjacking, DOM Manipulations, and Text Encoding) both before and after implementing interventions (such as parser modifications or adversarial training). We found that after applying adversarial patches, the True Positive Rate (TPR) for all attack categories, except BJE, improved beyond their initial training accuracies. For instance, regular phishing attacks achieved a TPR of 99.4\%, while DOM manipulations reached 98\%.

Despite the application of adversarial patches, websites with Behavioral JavaScript Evasion (BJE) continued to have a comparatively lower TPR (85.9\%) than other attack categories. Upon closer inspection, we discovered that a significant portion of these samples were affected by the ObfuscateJS (A8) attack. Given the low distribution of BJE samples in our ground truth dataset (n=715), we decided to collect more samples for this attack category. From July 11th to 29th, 2024, we gathered the source code for 17,305 new websites from PhishTank which had been verified as phishing. Using Esprima and heuristics highlighted in prior literature, we identified and manually verified 411 new BJE samples. These samples were then adversarially augmented using the optimization algorithm detailed in Section~\ref{adversarial_impact}. Both the original and adversarial samples were added to the ground truth and used for retraining. Following the retraining, we observed a significant reduction in the adversarial advantage of the BJE samples, both with and without the patch. Without the patch, the median adversarial advantage decreased to 0.084, and after applying Patch 5, it further reduced to 0.039, achieving a TPR rate of 97.2\%. In the next section, we identify whether the performance boost due to adversarial training transfers to real-world websites outside the ground-truth dataset.

We also re-evaluated our adversarially patched model on the metadata of the 42.7M domains collected from Certstream that was collected in Section~\ref{identifying_new_urls}. we find that 1,778 more websites were flagged as phishing,  a noticeable increase of 6.88\% compared to when the initial model was run (25,796 URLs to 27,574 URLs). Breaking down these numbers per category: The model detected 18,549 regular (6.6\% increase), 3,391 JavaScript evasion (7.3\% increase), 3,583 Clickjacking (7.0\% increase), 1,147 DOM (8.5\% increase), and 904 Text encoding attacks (8.3\% increase). However, it is more important to evaluate the False positive and False negative rates of these samples. We pseudo-randomly selected the same number of URLs (2.5k) from the detected set, with the same per category distribution, with 1,632 regular attacks, 280 BJEs, 313 clickjacking, 94 DOM Manipulations, and 78 text encoding. Similar to the evaluation of the initial model, we also randomly selected 2.5k benign websites. Table~\ref{table:model_evasive_attacks_after_patching} illustrates the performance of the model against different categories of attacks. We see a noticeable boost across all attack categories, for example, for Behavior-based JS evasion attacks, the model now achieves an accuracy of 94.8\% (an increase of 1.9\%), whereas for Text Encoding attacks, the accuracy increased to 95.7\% (an increase of 3.2\%). This indicates that the boost in detection for attack categories 
transferred from the test set (as seen in the previous section) to real-world websites.

\begin{table}[t]
\caption{Performance of PhishLang against evasive attacks after adversarial patching. Values in parenthesis denote \% increase}
\label{table:model_evasive_attacks_after_patching}
\centering
\resizebox{0.9\columnwidth}{!}{%
\begin{tabular}{l|c|c|c|c}
\hline
\textbf{Attack Type} & \textbf{Samples} & \textbf{Accuracy} & \textbf{Precision} & \textbf{Recall} \\ \hline \hline
Regular Attacks & 1,623 & 97.0\% (2.8\%) & 96.0\% (1.5\%) & 98.3\% (2.6\%) \\ 
Behavior-based JS & 280 & 94.2\% (1.3\%) & 95.6\% (2.3\%) & 94.1\% (1.5\%) \\
Clickjacking & 313 & 93.5\% (2.2\%) & 95.7\% (3.6\%) & 94.0\% (3.5\%) \\ 
DOM Manipulations & 94 & 94.1\% (5.7\%) & 94.8\% (5.8\%) & 93.5\% (6.4\%) \\ 
Text Encoding & 78 & 95.5\% (4.8\%) & 96.4\% (5.2\%) & 95.1\% (5.2\%) \\ \hline
\end{tabular}}

\end{table}

\subsection{Evaluating misdetections}
\label{evaluating_misdetections}

\begin{figure}[!htb]

\centerline{\includegraphics[width=1\columnwidth]{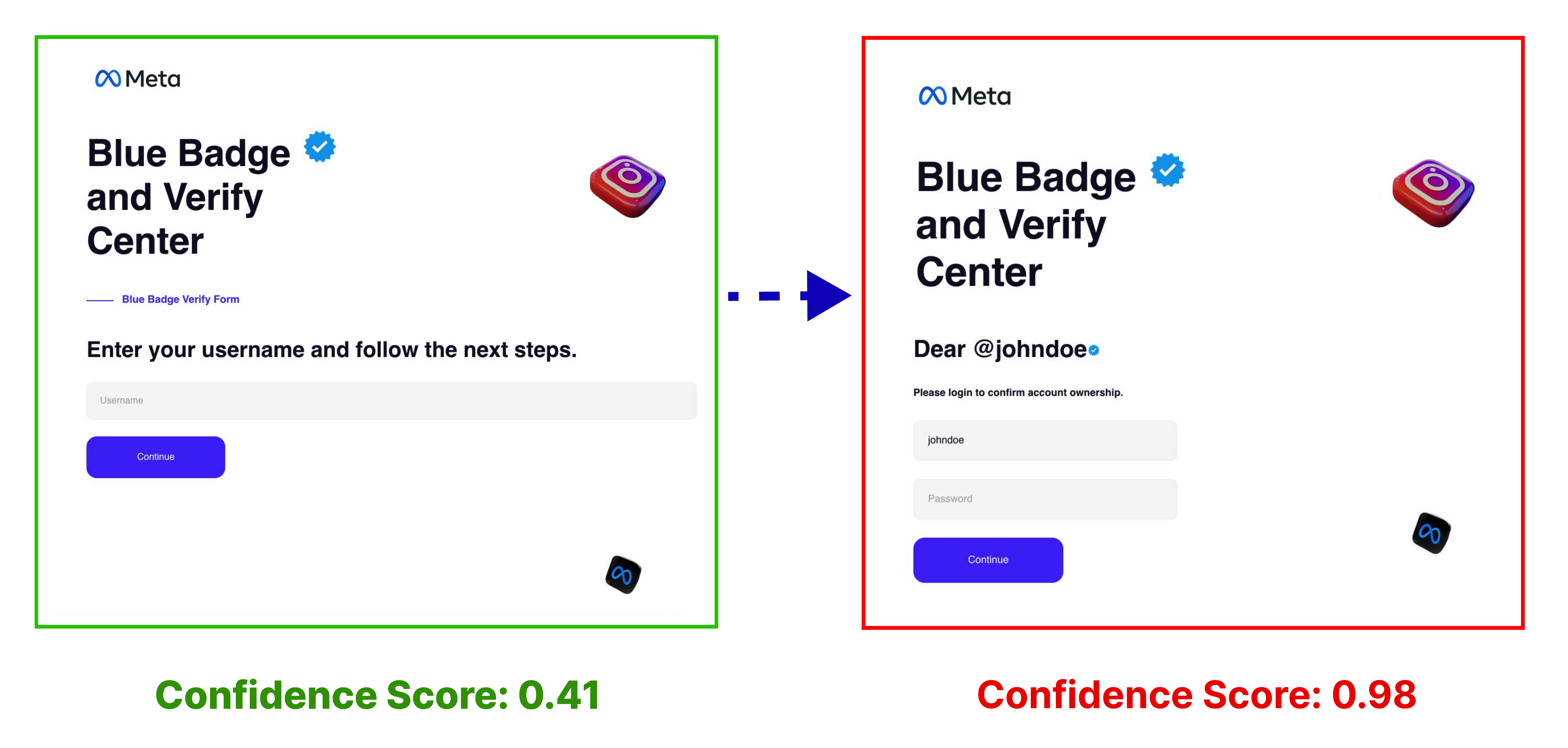}}
\caption{Example of an Instagram attack where the seemingly benign landing page (which was missed by PhishLang) led to a phishing page.}
\label{fig:delayed_detection}
\end{figure}

In this section, we evaluate the misdetections (False Positives and False Negatives) of PhishLang for both the vanilla model and after adversarial retraining on the 2.5K websites detected as phishing by PhishLang. We also analyzed the reasons for false positives and false negatives, observed whether adversarial improvements introduced additional false negatives, and monitored changes in the false positive and false negative rates for evasive attack categories post-adversarial training.


\subsubsection{False Negatives}

\begin{figure}
\centerline{\includegraphics[width=0.65\columnwidth]{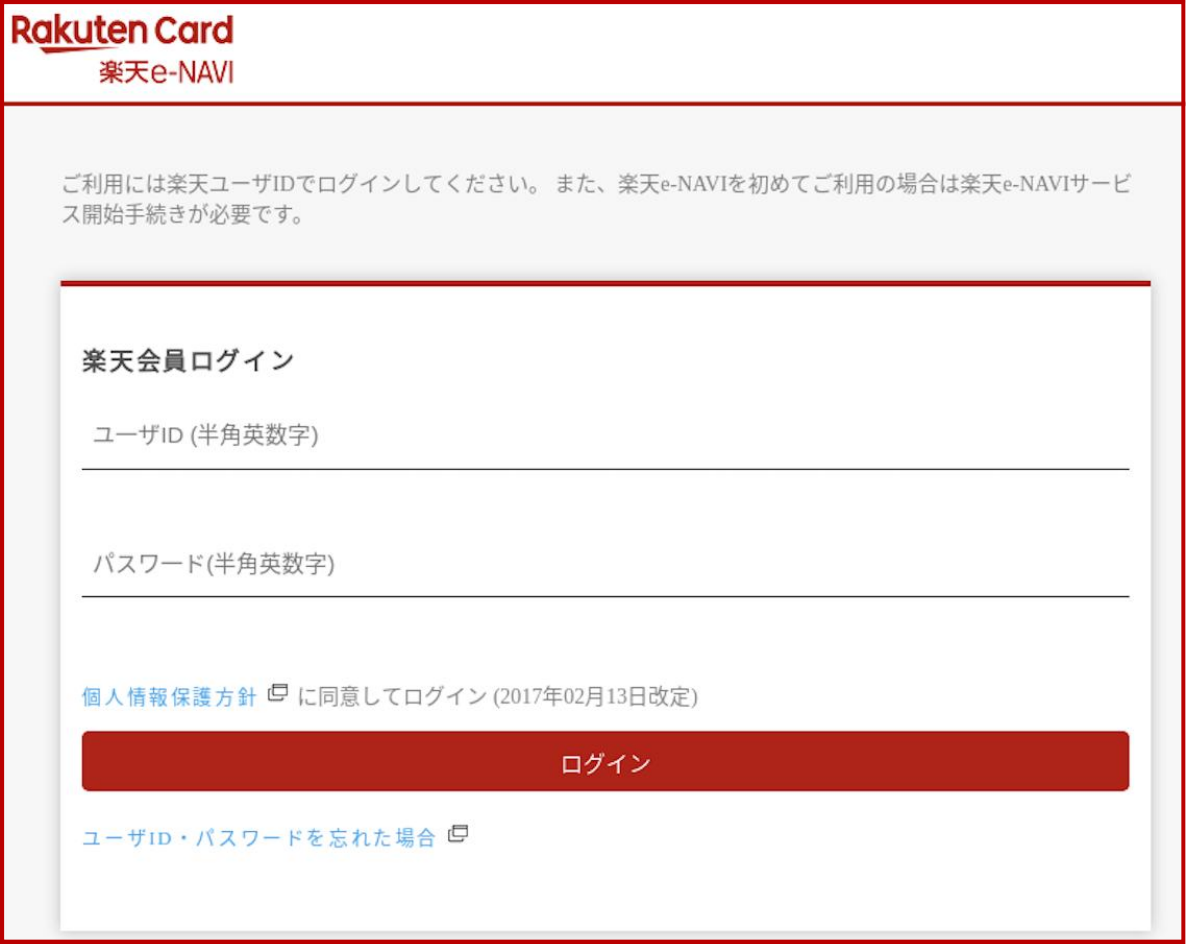}}
\caption{Example of a non-English (Japanese) false negative sample}
\label{fig:fn_1}
\end{figure}

Out of 261 undetected websites by PhishLang, 82 had CAPTCHAs and 36 had QR codes, both common phishing evasion tactics~\cite{vidas2013qrishing,kang2010captcha}. PhishLang successfully detected all phishing websites when QR code URLs were scanned and most CAPTCHA-protected sites, except two inactive ones. For QR codes, we used the \emph{pyzbar}\cite{pyzbar} library to automate scanning. However, solving CAPTCHAs programmatically can violate many websites' terms of service~\cite{sivakorn2016m}, so PhishLang detects phishing post-CAPTCHA.
We also found 25 sites that mimicked legitimate brands but initially only had input fields without malicious behavior. PhishLang's confidence was low on these landing pages, resulting in false negatives, but it detected phishing on the subsequent pages. Figure~\ref{fig:delayed_detection} shows an Instagram phishing page where PhishLang's confidence increased from 0.41 on the first page to 0.98 on the second. LIME analysis revealed PhishLang was influenced by brand-related text and form fields but lacked other phishing indicators initially. To address this, our framework now inputs placeholder data on benign-marked sites, submits it, and reassesses the destination, which helped detect all previously missed sites.
Additionally, 118 undetected websites were in non-English languages: Japanese (57), Spanish (36), German (18), and Italian (7). Figure~\ref{fig:fn_1} shows an example in Japanese. Using \emph{langdetect}, we found our training data was predominantly in English (n=18,683) with minimal coverage of these languages (Japanese=500, German=660, Spanish=403, and Italian=85). This lack of diversity likely caused misclassifications. While most phishing websites are in English, non-English phishing scams are emerging~\cite{darktrace2024lost}.
\textbf{Post-Adversarial Improvements:} Implementing the patches from Section~\ref{adversarial_impact} significantly improved performance. Only 2 out of 36 QR-code attacks, 5 out of 25 multi-page attacks, and 7 true-positive detections were misclassified as benign, despite a median confidence score of 0.44.

\subsubsection{False Positives}
\begin{figure}
\centerline{\includegraphics[width=0.6\columnwidth]{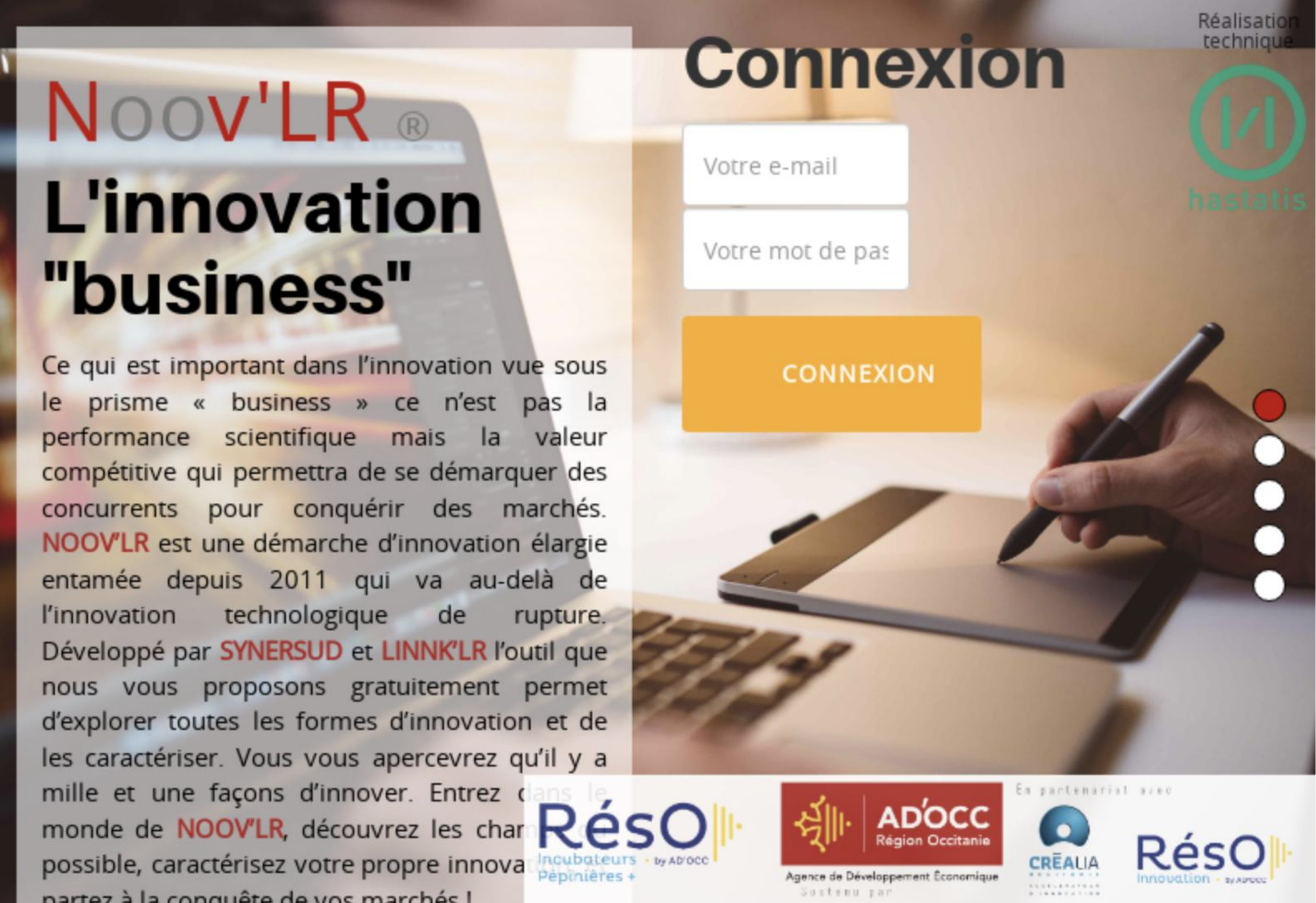}}
\caption{An example of a false positive with poor layout}
\label{fp_2}
\end{figure}

\begin{figure}[h!]
\centerline{\includegraphics[width=0.6\columnwidth]{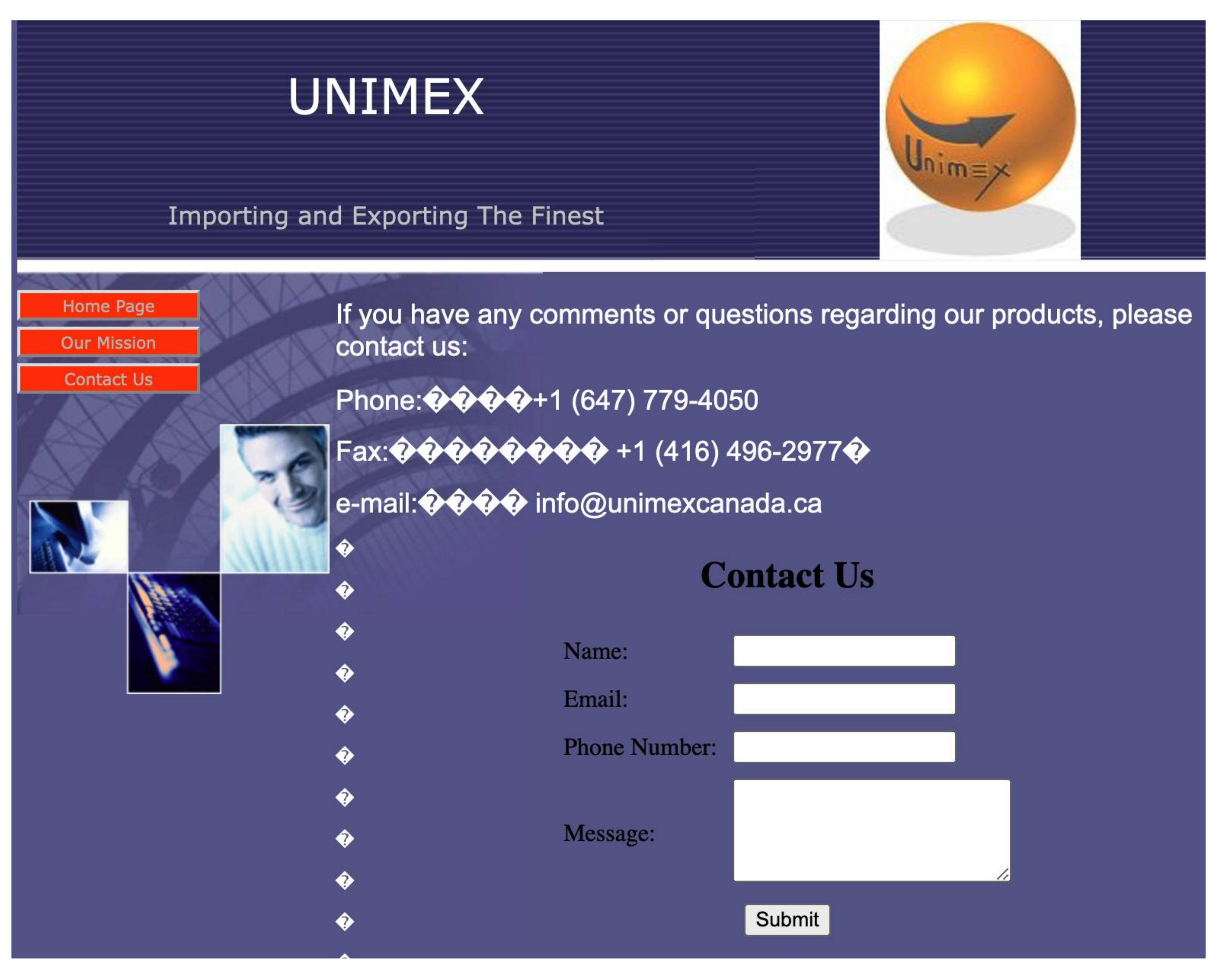}}
\caption{Example of a poorly designed (benign)website that was detected as a false positive }
\label{fig:fp_1}
\end{figure}

Upon evaluating the 112 false positives, we identified 42 malformed websites with poorly structured text and input fields soliciting sensitive information. For example, Fig ~\ref{fig:fp_1} shows a website with invalid characters and poorly designed input fields for personal information like name, email, and phone number, lacking clear instructions or context. These issues likely caused the site to resemble phishing samples, leading to incorrect classification by PhishLang.
Additionally, 23 websites featured multiple empty links and malformed meta tags with several redirections. Figure~\ref{fp_2}  illustrates an example where poorly rendered input fields and empty links indicated an incomplete population of the website. Addressing these false positives without compromising PhishLang's detection features is challenging.
We also discovered 14 websites with simple login fields that did not transmit data and contained no other information. These sites were classified as false positives because they neither imitated any organization nor had a deceptive motive, despite fitting traditional phishing profiles. Our crawler detects websites as soon as they appear on Certstream, so attackers could later add more details to these sites.

\begin{figure}
\centerline{\includegraphics[width=0.6\columnwidth]{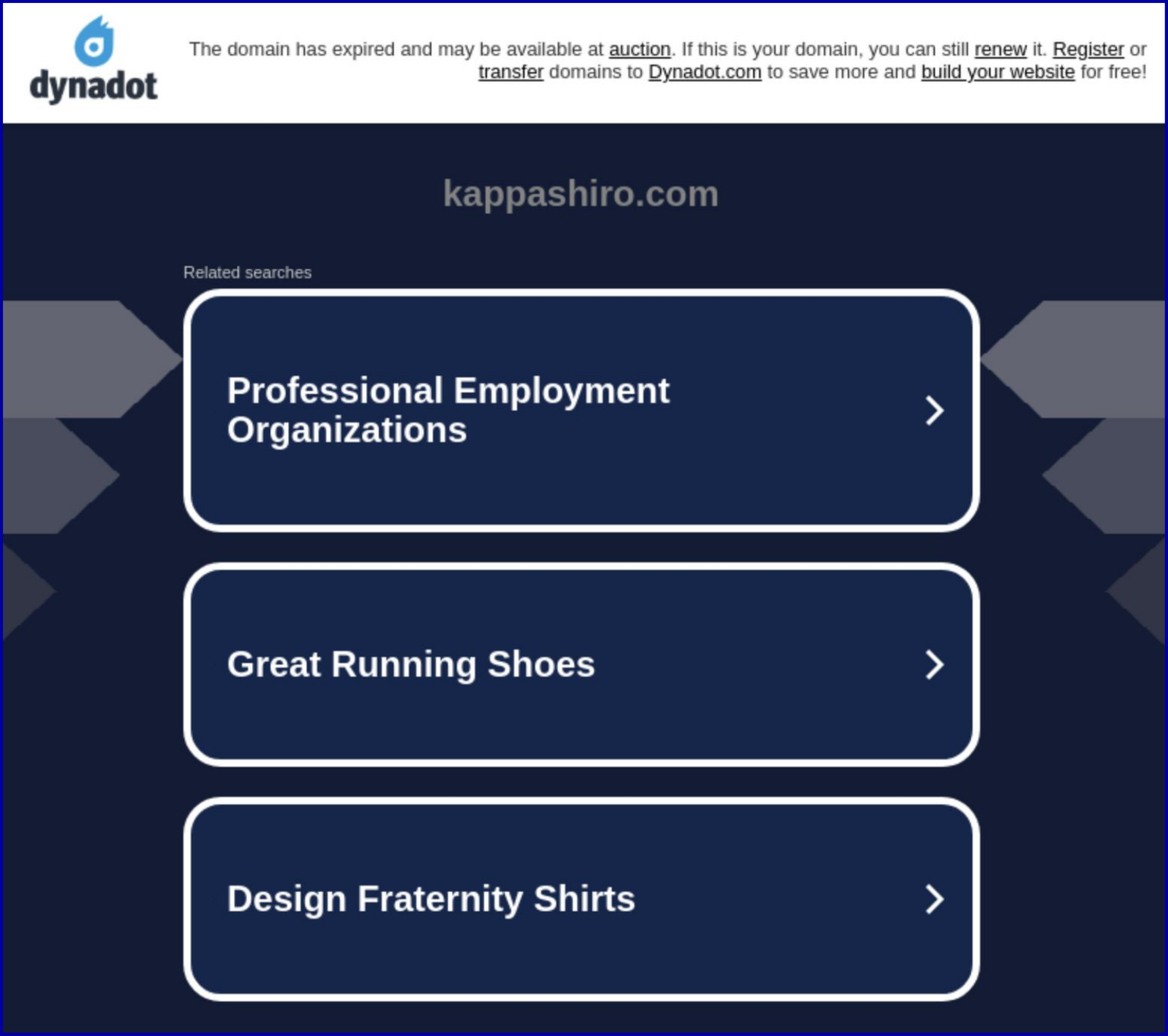}}
\caption{An example of a parked domain page that was flagged as a false positive}
\label{fp_3}
\end{figure}

We found 21 websites with domain parking pages, likely detected due to their lack of substantial content, presence of multiple advertising links, and frequent redirections. Figure~\ref{fp_3} shows one such example. To assess PhishLang's robustness against parked pages, we used keywords typical of parked pages to select 1,000 parked websites from our total sample of 42.7 million domains. PhishLang flagged only 5 of them as phishing (median confidence of 0.27).

Lastly, we noted 12 websites lacking noticeable phishing artifacts, with low confidence scores (Median=0.52). The small number of such samples might not significantly impact the training outcome if included in the dataset. We plan to retrain the model after gathering enough samples matching this criterion.
\newline
\textbf{Post-Adversarial Improvements:} Implementing adversarial improvements discussed in Section~\ref{adversarial_impact}, PhishLang correctly identified 28 of the websites with poor structural design as benign. Out of the 2.5k manually observed true positives, no new samples were changed from phishing to benign, increasing the true positive rate of our model (on the initial dataset) to 96.6\%.

\section{Client-side extension}
\label{client-side-extension-main}
One of the primary motivations behind designing PhishLang was to create a detection framework that is both efficient and resource-intensive, allowing users to run the protection measures entirely on the client side without relying on online blocklists. 
Thus we build a client-side application that integrates a Chromium-based browser extension with a background process, that can block phishing websites \textit{without} consulting any online blocklists. Once installed, the application automatically sets up our trained MobileBERT-based model as a local server, which the browser extension leverages to analyze all visited URLs. This setup is compatible with Microsoft Windows 10 and later versions, as well as Debian-based Linux distributions like Ubuntu (with MacOS support coming soon). The installation process configures the model, initiates the local service, and installs the extension for supported Chromium-based browsers. Operating entirely on CPU DRAM, the phishing detection processing is conducted entirely on the local machine. 

This approach offers several advantages. \textbf{Firstly,} it ensures that the user's browser history remains private and is not sent to any remote server. Prior studies have shown that users are often hesitant to share their browser history with third parties~\cite{mangio2020hands}. \textbf{Secondly,} as discussed in Section~\ref{blocklist_measurement}, numerous studies, including our own, have demonstrated that online anti-phishing solutions frequently suffer from delays and inefficiencies when detecting both standard and evasive phishing threats. These delays can be particularly problematic when such solutions are integrated into anti-phishing browser extensions, significantly slowing down threat detection. By using our extension, users benefit from a robust detection model running directly on their own systems, thereby avoiding the delays associated with blocklists.
\textbf{Thirdly,} as an open-source tool, our extension circumvents the need for substantial investment in server infrastructure, which would be necessary if detection were handled server-side. This setup not only limits scalability but also introduces risks of latency issues and downtimes. Running the extension entirely on the client side eliminates these concerns.

\begin{figure}[]
\centerline{\includegraphics[width=1\columnwidth]{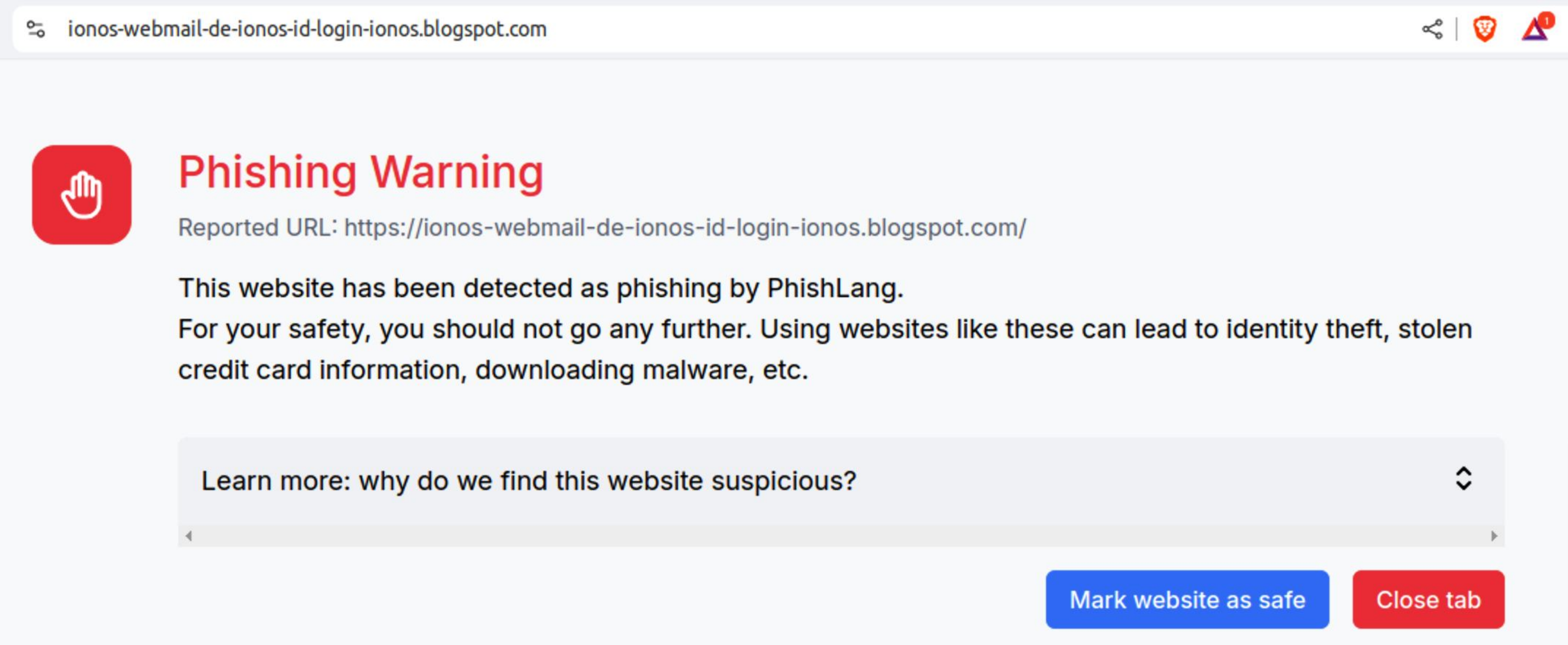}}
\caption{PhishLang client-side extension running in Brave Browser. Note that the website URL is not redacted since it is already inactive.}
\label{fig:phishlang_extension}
\end{figure}

Also, recent research~\cite{pourmohamad2024deep} has indicated that client-side implementations of in-browser anti-phishing measures, such as Google Safe Browsing, typically rely on simple heuristics and basic machine learning algorithms. These methods often lead to high false positive rates and low detection rates (14\%) for threats when not using online blocklists. In contrast, our extension offers an open-source alternative that utilizes a more advanced and flexible model architecture. In Section~\ref{clientside_performance}, we evaluate the performance and efficiency of our extension across various system configurations. Our findings indicate that the extension uses between 28MB and 247MB of RAM at peak load, which is a small fraction of the memory available in most modern systems. It can also run efficiently on low-end systems with only 1GB of RAM and a single-core processor while maintaining high precision (0.95) and recall (0.96) rates. To enhance performance, the application includes an optional built-in whitelist that users can choose to disable. Users also have the option to report URLs to our PhishLang database, though this feature is completely voluntary and does not prompt for user action. Additionally, users can add detected websites to a personal 'Safe List' or report URLs as false positives to the PhishLang database if necessary. Figure~\ref{fig:phishlang_extension} provides an illustration of the client-side extension in operation.

\subsection{Evaluating the performance of our client-side extension}
\label{clientside_performance}

We evaluate 2,000 verified phishing websites from PhishTank, comparing the performance of PhishLang on four distinct system configurations. The first configuration is our experimental setup, featuring an Intel Xeon W Processor with 184GB of RAM and 4x NVIDIA A5000 GPUs, running Ubuntu 22.04 LTS, with the model running natively. The second configuration is a mid-tier setup, consisting of an Intel Core i5 13th Gen processor, 8GB of RAM, no GPU, and running Windows 11 Build 22631. The third configuration is a low-tier setup, using an Intel Celeron N4500 processor, 2GB of RAM, no GPU, and running Ubuntu 24.04 LTS. The final configuration involves a virtual machine using VMware Player emulated with 2 core, 4GB of RAM, no GPU acceleration, and running Windows 11 Build 22631. These various configurations were chosen to assess PhishLang's performance across a spectrum of hardware configurations that reflect typical end-user environments.

Table~\ref{tab:phishlang_configs} presents the performance of PhishLang in terms of efficiency, memory usage, and inference time. Our findings show that the client-side application running on both Ubuntu and Windows 11 achieves the same efficiency as the model running on our experimental server, indicating no performance loss when PhishLang operates on the client-side. Additionally, all configurations exhibit similar memory usage. Memory usage was calculated by considering the system service (local server) and the memory consumed by the browser extension. It is noteworthy that the experimental setup has lower RAM usage since it runs the base model without any browser extension overhead. Regarding inference time, the low-tier and virtual machine configurations, despite having weaker processors, demonstrate only a slight decrease in speed compared to the experimental and mid-tier setups. All configurations maintain a median inference time of under one second. This suggests that our application, powered by the lightweight MobileBERT architecture, offers rapid inference regardless of the system configuration. It is important to note that the reported inference time excludes the time required to fetch the website, as server congestion or intentional delays by attackers can significantly affect website loading times.

\begin{table}[]

\resizebox{1\columnwidth}{!}{%
\begin{tabular}{c|c|ccc|ccc}
\hline
\multirow{2}{*}{Configuration} & \multirow{2}{*}{Precision/Recall} & \multicolumn{3}{c|}{Memory usage} & \multicolumn{3}{c}{Inference time (in secs)} \\ \cline{3-8} 
 &  & \multicolumn{1}{c|}{Min} & \multicolumn{1}{c|}{Max} & Median & \multicolumn{1}{c|}{Min} & \multicolumn{1}{c|}{Max} & Median \\ \hline
Experimental & 0.95/0.96 & \multicolumn{1}{c|}{28MB} & \multicolumn{1}{c|}{215MB} & 76MB & \multicolumn{1}{c|}{0.08} & \multicolumn{1}{c|}{1.09} & 0.44 \\ \hline
Medium-end & 0.95/0.96 & \multicolumn{1}{c|}{51MB} & \multicolumn{1}{c|}{308MB} & 93MB & \multicolumn{1}{c|}{0.13} & \multicolumn{1}{c|}{1.47} & 0.59 \\ \hline
Low-end & 0.95/0.96 & \multicolumn{1}{c|}{54MB} & \multicolumn{1}{c|}{247MB} & 89MB & \multicolumn{1}{c|}{0.18} & \multicolumn{1}{c|}{1.95} & 0.73 \\ \hline
Virtual Machine & 0.95/0.96 & \multicolumn{1}{c|}{68MB} & \multicolumn{1}{c|}{231MB} & 102MB & \multicolumn{1}{c|}{0.11} & \multicolumn{1}{c|}{2.43} & 0.81 \\ \hline
\end{tabular}}
\caption{Performance of the PhishLang app for various system configurations
}
\label{tab:phishlang_configs}
\end{table}

\subsubsection*{Trade-offs:} Currently, running PhishLang as a client-side application requires downloading TensorFlow and PyTorch, which the installer automatically handles if they are not already installed. This setup can take up to 2.5GB of storage space on the user's machine. However, since storage has become significantly cheaper over the years, it is reasonable to assume that most end-user systems can accommodate these requirements.
An alternative to reduce storage space is to implement the model using TensorFlowJS directly within the browser extension. However, this approach has significant drawbacks. Running the model in the browser with TensorFlowJS may lead to slower performance compared to running it natively as a local server application, as browser-based implementations (especially on lower-end configurations) do not leverage hardware acceleration as effectively. Additionally, browser environments can be inconsistent across various devices and operating systems, which might lead to compatibility and stability issues. This inconsistency could result in the extension not functioning uniformly across all browsers, potentially affecting the reliability of the phishing detection system. Moreover, implementing the extension separately for different browsers would add to the development complexity. In contrast, a local server-based implementation would ensure broader compatibility, as it only requires basic communication capabilities between the extension and the local server, a feature supported by all major browsers.

\section{Conclusion}
\label{conclusion}
In this work, we introduced PhishLang, a novel phishing detection framework leveraging the MobileBERT architecture to identify phishing websites by comparing source codes with established phishing patterns. PhishLang outperforms various machine learning and deep learning models, reducing detection time and resources while showing strong resilience against adversarial attacks. In real-time, it identified several new phishing websites, including evasive attacks that were missed by popular anti-phishing tools. Reporting these threats improved detection rates, demonstrating PhishLang's impact. We also implemented PhishLang as the first open-source client-side browser extension that does not require an external blocklist.

\bibliographystyle{ACM-Reference-Format}
\bibliography{main}

\end{document}